\documentstyle[aps,prl,twocolumn,graphicx,epsfig,amssymb]{revtex}

\begin{document}


\twocolumn[\hsize\textwidth\columnwidth\hsize\csname@twocolumnfalse\endcsname

\title{Anomalous Superconductivity and Field-Induced Magnetism in CeCoIn$_{5}$}

\author{T. P. Murphy, Donavan Hall, E. C. Palm, S. W. Tozer, C.
Petrovic$^{\dag}$, Z. Fisk}
\address{National High Magnetic Field Laboratory, Florida State University,
Tallahassee, FL 32310}

\author {R. G. Goodrich}
\address{Department of Physics and Astronomy, Louisiana State
University, Baton Rouge, LA 70803-4001}

\author{P. G. Pagliuso, J. L. Sarrao, J. D. Thompson}
\address{Los Alamos National Laboratory, Los Alamos, NM  87545}

\date{\today}

\maketitle

\begin{abstract}
In the heavy fermion superconductor CeCoIn$_{5}$ (T$_{c}$=2.3K)
the critical field is large, anisotropic and displays hysteresis.
The magnitude of the critical-field anisotropy in the a-c plane
can be as large as 70 kOe and depends on orientation. Critical
field measurements in the (110) plane suggest 2D
superconductivity, whereas conventional effective mass anisotropy
is observed in the (100) plane. Two distinct field-induced
magnetic phases are observed: H$_{a}$ appears deep in the
superconducting phase, while H$_{b}$ intersects Hc$_2$ at T=1.4 K
and extends well above T$_c$. These observations suggest the
possible realization of a direct transition from ferromagnetism to
Fulde-Ferrel-Larkin-Ovchinnikov superconductivity in CeCoIn$_{5}$.
\end{abstract}
\pacs{74.25.Dw,74.25.Ha,74.70.Tx,75.30.Gw}
]

% body of paper here
%\section{Introduction}

The interaction of magnetism and superconductivity is a
significant and long standing problem in condensed matter physics.
Usually, the presence of magnetic order undermines
superconductivity, but in heavy fermion materials,
superconductivity and magnetism can coexist without deleterious
consequences to the superconducting state. These systems provide
an opportunity to explore the interaction of magnetic and
superconducting order parameters as a function of temperature,
pressure, or magnetic field\cite{Mathur}. While antiferromagnetism
interacting/coexisting with superconductivity is the case most
often considered, examples of ferromagnetism coexisting with
superconductivity have been reported
recently\cite{uge2_ggl,zrzn2_ggl}. In heavy fermion
superconductors the combination of large initial critical field
vs. temperature slopes and long mean free paths also potentially
allows for the observation of critical fields beyond the Pauli
limit and, perhaps, inhomogeneous pairing states
\cite{Gloos,fflo}.

Recently the heavy fermion compound CeCoIn$_{5}$ was observed to
superconduct at 2.3 K, the highest T$_c$ yet reported for a heavy
fermion superconductor\cite{Petrovic2000}.  Specific heat and
thermal transport studies establish that the superconductivity in
this material is unconventional and magnetically
mediated\cite{Movshovich2001}. Because crystallographic anisotropy
might play an important role in the properties of this tetragonal
material\cite{Petrovic2000} and de Haas-van Alphen measurements
reveal a two-dimensional character of the Fermi
surface\cite{Hall2001b,settai}, a thorough investigation of the
anisotropic critical field-temperature phase diagram was
undertaken and is reported in this Letter.  We observe not only an
upper critical field, Hc$_2$, that varies differently as a
function of angle in the (100) and (110) planes but also the
existence of field-induced magnetic phases in both the normal and
superconducting states of CeCoIn$_{5}$.

%\section{Experiment}

CeCoIn$_{5}$ forms in the tetragonal HoCoGa$_{5}$ crystal
structure with lattice constants a=4.62$\AA$ and c=7.56 $\AA$
\cite{Petrovic2000,Kalychak}. The crystal structure consists of
alternating layers of CeIn$_{3}$ and CoIn$_{2}$. The
crystallographic axes of the flux-grown single crystals used in
our experiments were determined by Laue x-ray diffraction. The
[001] axis was parallel to the shortest dimension of the crystal
and [100] and [010] axes were parallel to the natural
edges of the nominally square crystals. The superconducting-normal
phase boundary and the magnetization of CeCoIn$_{5}$ were
determined by electrical transport, AC susceptibility and
cantilever magnetometery measurements as a function of magnetic
field (0-200 kOe) and temperature (0.020-27 K). Angular variations
were measured using a rotating sample stage\cite{Palm} in a
top-loading dilution refrigerator and in a $^3$He cryostat. Three
different single crystals were studied, with consistent agreement
among their measured Hc$_2$ values.

\begin{center}
\begin{figure}[tbp]
    \includegraphics[scale=0.45]{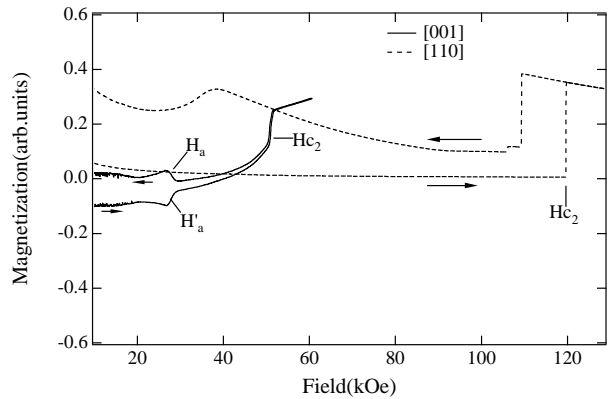}
    \caption{Magnetization loops for field applied along [110] and [001] 
    at 20 mK in CeCoIn$_5$. Arrows indicate direction of field sweep.}
\label{Fig1}
\end{figure}
\end{center}

%\section{Results}
Fig. \ref{Fig1} shows a signal proportional to $\vec{M}$ plotted
against applied magnetic field for both increasing and decreasing
fields. The two traces are for the field applied along the [110]
and [001] crystal axes at T = 20 mK.  With increasing field, a
narrow superconducting-normal transition,
$\Delta$H$c_{2}$$\textless$10 Oe, is clearly seen in the
$\vec{H}\parallel$[110] trace, and a somewhat broader transition
is seen in the $\vec{H}\parallel$[001] trace. These traces are
typical of data used to construct the phase diagrams reported
below. At T=20 mK and $\vec{H}\parallel$[001], Hc$_{2}$ =50.5 kOe
and for $\vec{H}\parallel$[110], Hc$_{2}$ =119 kOe. Resistivity
measurements (not shown) confirm that these transitions correspond
to superconducting-normal transitions. For a material having
T$_c$=2.3 K, these values of Hc$_2$ are quite large:  a simple
estimate of the Clogston limit gives Hc$_{2}$(T=0) = (18.6
kOe/K)T$_{c}$ = 43 kOe \cite{Clogston}. For field applied along
[110] the normal-superconducting resistive transition occurs at
the same field on both upsweep and downsweep; however, the
magnetization transition occurs at a lower field on the downsweep,
suggesting an additional phase transition in the superconducting
state. No such second transition is observed for H $\parallel$
[001].

 An additional feature apparent in Fig.
\ref{Fig1} is the peak in magnetization observed for H $\parallel$
[001] at $H_{a}$=28 kOe. Preliminary investigations show that this
feature appears only below 100 mK and exhibits a complex
dependence on field orientation and sweep direction.  Although it
will be discussed in detail elsewhere\cite{murphy}, we note here
that $H_{a}$ appears to merge with Hc$_2$ (upsweep) when the
applied field is within 5 degrees of [110].

\begin{center}
\begin{figure}[tbp]
    \includegraphics[scale=0.45]{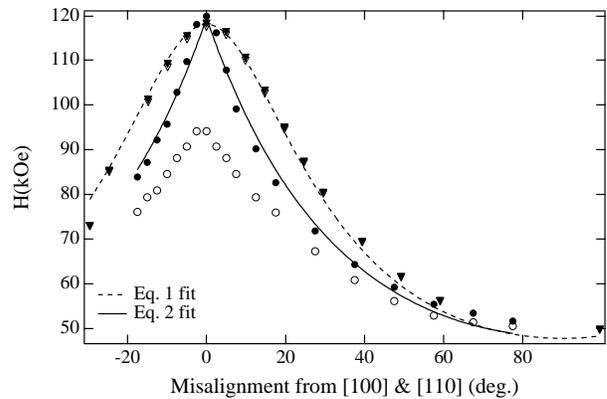}
    \caption{Hc$_{2}$ as a function of angle for CeCoIn$_{5}$
    for field rotations from [110] to [001] ($\bullet$= 
    upsweep,$\circ$= downsweep) and from [100] to [001] 
    ($\blacktriangledown$= upsweep, $\triangledown$=
    downsweep).The data points for the second curve fall on top of 
    one another thus the open triangles are not visible. See text for fit equations. }
\label{Fig2}
\end{figure}
\end{center}

The angular dependence of Hc$_{2}$ at 20 mK is shown in
Fig.\ref{Fig2}. The evolution of Hc$_{2}$ for rotation of
$\vec{H}$ from [001] into [100] is well described by the
anisotropic effective mass model\cite{levb}, taking
Hc$_{2}$($\theta$) as the upsweep value:
\begin{equation}
    Hc_{2}(\theta)=Hc_{2}(\theta=0)/[cos^{2}(\theta) + \alpha sin^{2}(\theta)]^{1/2}
    \label{eq:massvar}
\end{equation}
where $\theta$ is the angle of the applied field out of the
tetragonal basal plane and $\alpha$ is the ratio of effective
masses m*($\theta=0$)/m*($\theta=90$).  The large value of
$\alpha$=6.1 confirms the significant electronic anisotropy in
CeCoIn$_5$ deduced from de Haas van Alphen
measurements\cite{Hall2001b,settai}.

Rotating $\vec{H}$ from [001] into [110] produces a much more
cusp-like angular dependence than Eq. \ref{eq:massvar} would
predict. In this case, the data are well described by Tinkham's
equation for Hc$_{2}$ as a function of angle in thin film
superconductors\cite{Tinkham}:

\begin{equation}
    |Hc_{2}(\theta)sin(\theta)/Hc_{2}(90)| +[Hc_{2}(\theta)cos(\theta)/Hc_{2}(0)]^{2}=1.
    \label{eq:thinfilm}
\end{equation}

Both sets of data in Fig. \ref{Fig2} were obtained using the same
single crystal, so neither sample-to-sample variation nor
demagnetization corrections can explain the different angular
variations in Hc$_{2}$. We also note that Hc$_{2}$[110]=119 kOe
while Hc$_{2}$[100]=118 kOe which implies the existence of
in-plane anisotropy in Hc$_{2}$\cite{murphy,izawa}.

The angular dependence of Hc$_{2}$ observed in the (110) plane is
reminiscent of behaviors in granular thin film\cite{Wu,Adams} and
multilayer\cite{Ogrin} systems. In fact the quality of the fit to
our data is comparable to and extends over a wider angular range
than that in Al films\cite{Wu}. Why 2D behavior in one particular
plane would be observed in bulk CeCoIn$_{5}$ is not understood.
Band structure calculations suggest that the density of states in
the MIn$_{2}$ layer in CeMIn$_{5}$ is quite low\cite{haga} and
leads to the speculation that perhaps CeCoIn$_{5}$ may behave as a
pseudo CeIn$_{3}$:CoIn$_{2}$ multilayer system. Even if such a
speculation were shown to be relevant, why the phenomenon would
manifest itself in [001]-[110] rotations but not [001]-[100]
rotations is unclear; however, it might be related to an in-plane
modulation of the superconducting gap
function\cite{Movshovich2001,izawa} or to anisotropic Fermi
surface nesting\cite{Hall2001b,settai}.

The difference in field between the upsweep and downsweep
transitions in magnetization is a strong function of
crystallographic direction (Fig. \ref{Fig2}). The difference
increases as the field is rotated toward [110] and has a maximum
value of 25 kOe. As will be discussed below, CeCoIn$_5$ displays a
metamagnetic transition at high fields, and the presence of a
static magnetization in the sample complicates the determination
of the downsweep transition field. The values shown in Fig.
\ref{Fig2} have been corrected to account for the magnetization,
$\vec{M}$, in the sample that contributes to the internal magnetic
field, $\vec{B}$ according to the relation:
$\vec{B}=\vec{H}+\mu_0\vec{M}$. Using this relation, we corrected
an offset in the measured downsweep transition field around [110]
that was the result of moving from a magnetic normal state into a
superconducting one. The maximum contribution of $\vec{M}$ is
estimated to be $\mu_0\vec{M}$=15 kOe along [110]. The separation
in transition fields as a function of angle for the [100] rotation
is shown in Fig. \ref{Fig2} as well. The difference between the up
and down sweeps in this case is almost negligible (resulting in
symbols in the figure that essentially overlap). The maximum field
separation along [100] is only 0.8 kOe, a factor of 31 less than
the value of 25 kOe that is found along [110].

\begin{center}
\begin{figure}[tbp]
    \includegraphics[scale=0.45]{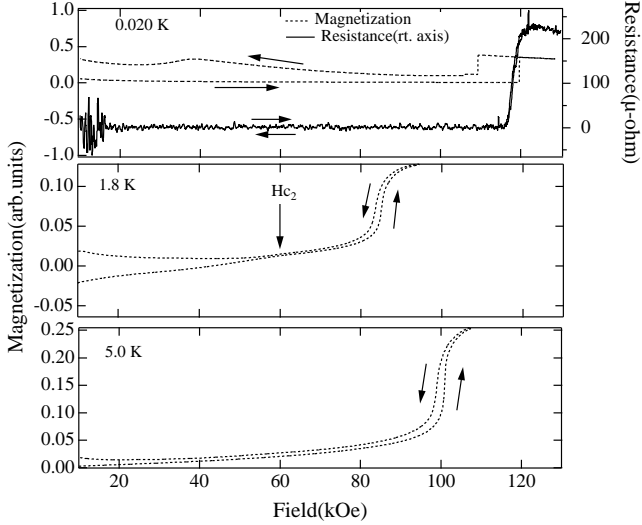}
    \caption{Magnetization as a function of field ($\vec{H}\parallel$ [110]) at 3
characteristic temperatures (0.020 K, 1.8 K, and 5 K) in CeCoIn$_5$. 
The noise in the 0.020 K resistance measurement is due to flux 
popping in the magnet at low fields. Note that the sharp transition 
in the magnetization in the 0.020K panel occurs at the onset of 
superconductivity as displayed in the resistance measurement. Arrow indicates position of zero-resistance transition for 1.8 K panel. }
\label{Fig3}
\end{figure}
\end{center}

\begin{center}
\begin{figure}[tbp]
    \includegraphics[scale=0.45]{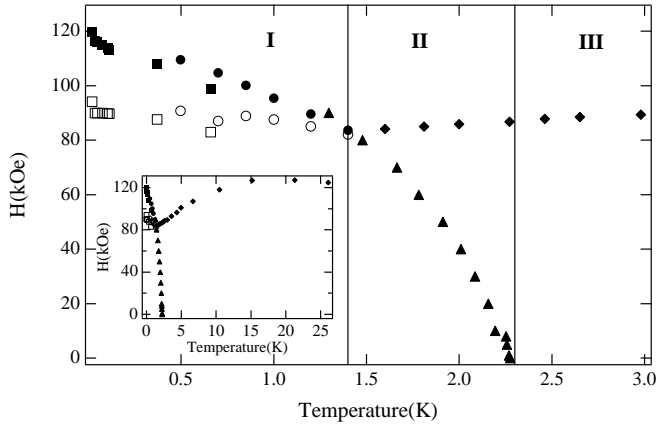}
    \caption{H-T diagram for CeCoIn$_{5}$ with $\vec{H}$ applied in the 
    (110) direction (the inset emphasizes the high-temperature range of 
    the main figure).Circles and squares denote  magnetization transitions 
    ($\blacksquare$=up sweep, $\square$=downsweep)and ($\bullet$=up 
    sweep, $\circ$= down sweep). Triangles indicate resistively determined Hc$_2$.
    Measurements were made with three different systems and the resultant
    offset between circles and crosses is due to  slight differences in crystal alignment.
    The diamonds denote the field-induced magnetic transition that appears
    at 1.4K. \textbf{I}, \textbf{II}, and \textbf{III} indicate the three 
    regions in the phase diagram discussed in the text. }
\label{Fig4}
\end{figure}
\end{center}

The evolution of these transitions in the [110] direction in
CeCoIn$_5$ with temperature also is anomalous. An H-T phase
diagram for H $\parallel$ [110] in CeCoIn$_5$ is shown in
Fig.\ref{Fig4}.  Three characteristic temperature ranges can be
identified (see Fig. \ref{Fig3} for representative data): I) For T
$<$ 1.4 K, field-separated transitions in magnetization are
observed and the changes in magnetization at the transitions are
step-like; II) for 1.4 K $<$ T $<$ 2.3 K, no evidence for the
lower-field transition nor step in magnetization at Hc$_2$ is
observed, but for fields greater than Hc$_2$, a normal-state
metamagnetic transition, occurring at H$_b$, is seen; and III) for
T $>$ 2.3 K only the metamagnetic transition is observed.

In Region I the field separation between magnetization transitions
decreases with increasing temperature and at 0.5 K,
Hc$_{2}(\theta)$ has the same relative angular dependence (not
shown) as at 20 mK (Fig. \ref{Fig2}). The zero-resistance
transition occurs (regardless of field sweep direction) at the
higher up-sweep value of Hc$_2$ deduced from magnetization, and a
new ferromagnetic-like (because of the observed steps in
magnetization) first-order transition in the superconducting state
emerges below the resistively-determined Hc$_2$. This state is
found below 0.6T$_{c}$ at high fields and its appearance depends
strongly on the orientation of field with respect to
crystallographic axes. Taken together, these observations are
consistent with a spatially inhomogeneous
Fulde-Ferrel-Larkin-Ovchinnikov (FFLO) state\cite{fflo}.  The fact
that no signature of a BCS-FFLO transition prior to the
superconducting-normal transition is observed in upsweep
magnetization may suggest that this transition is hysteretic in
field or that the FFLO state is only stabilized by the presence of
magnetic order. High-field heat capacity and neutron scattering
measurements should be able to clarify this issue.

Although the FFLO state is rarely observed\cite{norman,tenya},
CeCoIn$_{5}$ satisfies the essential conditions for its
existence\cite{singleton}:  it is in the clean
limit\cite{Movshovich2001}, has a quasi-2D Fermi
surface\cite{Hall2001b,settai}, and has an Hc$_{2}$ much larger
than the Clogston limit. The transition from the normal state to
the FFLO state is from ferromagnetic to superconducting, which to
the best of our knowledge, is unprecedented. Recent calculations
of Zeeman effects in d-wave superconductors\cite{yang} (e.g.,
CeCoIn$_{5}$\cite{Movshovich2001,izawa}) suggest that an increase
in Hc$_{2}$ and the appearance of another magnetic transition,
perhaps related to H$_{a}$, at lowest temperatures is a
consequence of an FFLO state in such a superconductor.  If we are
not observing FFLO superconductivity in the [110] direction, then
the finite jump in $\vec{M}$ in the superconducting state implies
the coexistence of superconductivity with a spin-polarized state,
the field-sweep dependent continuation of H$_b$(T) into the mixed
state.

The signature for Hc$_{2}$(T) intersects H$_{b}$(T) and the
signature of the FFLO state vanishes at H=80 kOe and T=1.4 K.
Because the magnetization change at Hc$_{2}$ disappears above 1.4
K , we used transport measurements to follow Hc$_{2}$ up to
T$_{c}$(H=0) with no observable hysteresis nor second transitions
present. The change in magnetization at H$_{b}$ is approximately a
factor of 2.5 less than at Hc$_{2}$ (for T $<$ 1.4 K) and appears
to be second order as a function of temperature. The signature for
H$_{b}$ weakens as $\vec{H}$ is rotated away from [110] and is
completely absent for H$\parallel$ [001], again illustrating the
anisotropic magnetic behavior of this material even in the normal
state. Given the extent to which its evolution is influenced by
superconductivity without deleterious effects on Hc$_2$, it is
tempting to identify the paramagnetic-magnetic normal state
transition with spin polarization of a sheet of Fermi surface;
quantum oscillation measurements to test this hypothesis are in
progress.

In summary, we find a remarkable Hc$_{2}$ anisotropy in
CeCoIn$_{5}$ that is correlated with the presence of a magnetic
transition in the superconducting state for H $\parallel$ [110].
These data can be described empirically in terms of 2D
superconductivity and suggest the formation of an FFLO state.
Hc$_{2}$ anisotropy also exists within the (100) plane but is
describable by anisotropic band structure effects, and does not
present hysteresis. We also have observed two new magnetic phases
in CeCoIn$_{5}$, one occurring deep in the superconducting state
at very low temperature (H$_a$) and the other (H$_b$) manifesting
itself as a field-induced metamagnetic transition that persists to
at least 25 K.

%\section{Acknowledgements}
We thank L.N. Bulaevskii, L.P. Gor'kov, M.P. Maley, and J. Singleton for fruitful discussions.
This work was performed at the National High
Magnetic Field Laboratory, which is supported by NSF cooperative agreement
No.  DMR-9527035, by the State of Florida, and Grant No. DMR-9971348
(Z.F.). Work at Los Alamos was performed under the auspices of the US Department of Energy.

% now the references. delete or change fake bibitem. delete next three
%   lines and directly read in your .bbl file if you use bibtex.

\end{document}